\documentclass[%
 reprint,
 superscriptaddress,
 amsmath,amssymb,
 aps,
 pra,
floatfix,
]{revtex4-2}

\newenvironment{captivy}[1]{
  \begin{tikzpicture}[every node/.style={inner sep=0}]
    \node[anchor=south west,inner sep=0] (image) at (0,0) {#1};
    \begin{scope}[x={(image.south east)},y={(image.north west)}]
}%
{
        \end{scope}%
  \pgfresetboundingbox
  \path[use as bounding box] (image.south west) rectangle (image.north east);
  \end{tikzpicture}%
}

\newcommand*{\oversubcaption}[3]{%
  \draw (#1) node[fill=white,inner sep=0pt, opacity=0.2, above, yscale=1.1, xscale=1.1] {\phantom{(a)#2}};
  \draw (#1) node[inner sep=0pt, above]{%
    \subfloat[#2\label{#3}]{\phantom{(a)}}
  };
}

\usepackage{hyperref}
\usepackage{graphicx}
\usepackage{dcolumn}
\usepackage{bm}
\usepackage{amsmath}
\usepackage{physics}
\usepackage{mathtools}
\usepackage{tikz}
\usepackage[caption=false]{subfig}
\usepackage[mathlines]{lineno}
\DeclareUnicodeCharacter{2212}{-}
\usepackage{enumitem}
\usepackage[export]{adjustbox}

\begin{document}

\preprint{APS/123-QED}

\title{Inducing Heat Reversal in a Three-Qubit Spin Chain}

\author{Saleh Naghdi}
 \email{mnaghdi@student.unimelb.edu.au}
\affiliation{%
 School of Physics, The University of Melbourne, Parkville, Victoria 3010, Australia
}%
\author{Thomas Quella}
 \email{thomas.quella@unimelb.edu.au}
\affiliation{%
 School of Mathematics and Statistics, The University of Melbourne, Parkville, Victoria 3010, Australia
}%
\author{Charles D.\ Hill}%
 \email{cdhill@unimelb.edu.au}
\affiliation{%
 School of Physics, The University of Melbourne, Parkville, Victoria 3010, Australia
}%
\affiliation{%
 School of Mathematics and Statistics, The University of Melbourne, Parkville, Victoria 3010, Australia
}%

\date{\today}

\begin{abstract}
By the standard second law of thermodynamics, heat spontaneously flows from a hotter body to a colder body. However, quantum systems in which quantum correlations play a prominent role can exhibit a non-classical reversal of such heat flow. We propose a quantum system consisting of a chain of qubits, each in local Gibbs states, where only adjacent qubits are allowed to thermally interact. By controlling initial quantum correlations along the chain, we then demonstrate non-classical heat reversal for the special case of a three-qubit chain on a quantum computer. We explore multiple initial conditions for the spin chain to showcase exotic behaviour such as the preferential pumping of heat afforded by unequal initial correlations between adjacent pairs of qubits, reinforcing the role that initial correlations play in influencing the dynamics of heat flow. 
\end{abstract}

\maketitle


\section{\label{sec:intro} Introduction}
Recent strides in quantum thermodynamics and quantum information science have discussed the role and utility of entanglement in giving rise to exotic thermodynamic behaviours \cite{resource} as it relates to entropy production \cite{entropy}, and anomalous equilibration \cite{asymmetric}. This has consequently led to a variety of applications for quantum thermal machines and quantum heat engines \cite{otto} by using quantum coherence to control heat \cite{nano} and other thermodynamic properties \cite{mutualinformation} on a microscopic scale.

One result obtained from the work done by Partovi \cite{partovi} challenges the classical notion that heat spontaneously flows from hotter to colder bodies -- famously referred to as the thermodynamic arrow of time \cite{schroeder_2021}. By proposing a particular entangled state between two multipartite quantum systems, Partovi shows that the anomalous reversal of heat flow from colder to hotter bodies can occur between systems that are initially quantum correlated. While peculiar in nature, this reversal effect emerges as a direct consequence of the quantum generalisation of the second law of thermodynamics which insists the change in the joint entropy of two systems is always non-negative. 

This idea was investigated more concretely by Jennings and Rudolph \cite{jennings} who examined the case of a three-qubit chain by way of simulations, using the initial condition in which the endmost qubits are correlated to showcase heat flow against the temperature gradient. Their work emphasises the complexity that initial correlations introduce to the flow of heat in a closed system. While Partovi, Jennings and Rudolph focused on a theoretical understanding of this phenomenon, it was not until recently that the reversal effect was experimentally observed by Micadei et al.\ for the simplest case of a pair of quantum correlated (as distinct from entangled) qubits prepared in a solution, using Nuclear Magnetic Resonance (NMR) spectroscopy \cite{micadei}. It is desirable to understand how more complex patterns of non-classical heat flow may arise by extending this setup to larger systems with more constituents.

In recent years, as noisy intermediate-scale quantum computers have emerged as a powerful device for simulating small-scale experiments, the field of quantum chemistry has enjoyed numerous advances in different areas, such as calculating the dipole moment of LiH in lithium-ion batteries \cite{lithium}, or learning the Hamiltonian of a quantum many-body system which has applications for statistical machine learning \cite{soleiman}. In this paper we harness the power of the IBM quantum computers to demonstrate the occurrence of heat reversal for the case of a quantum correlated three-qubit spin chain. By utilising the greater variety in initial conditions afforded by the jump to three qubits, we particularly demonstrate four main cases (Table \ref{tab:cases}) which illustrate and contrast the effects of quantum correlations with the classical, uncorrelated case. These cases reveal the exotic phenomena that quantum correlations introduce to heat flow, such as full and partial heat reversal, and the preferential pumping of heat in an initially symmetric temperature profile. Our system is similar to the one analysed in Ref.~\onlinecite{jennings} although, like Micadei et al., we additionally impose the condition that the qubits remain in a Gibbs state with respect to a fixed local Hamiltonian at all times so that it is possible to assign a temperature to each qubit at each instance in time. 


In Section~\ref{sec:theory}, we provide an outline of
the phenomenon of heat reversal with reference to quantum thermodynamic principles. We then define the initial state and interaction dynamics of our system consisting of an arbitrarily long chain of qubits, and in Section~\ref{sec:method} we detail one possible quantum circuit implementation for simulating the heat flow of the special case of a three-qubit chain on a quantum computer. The results of this heat flow for each of the four initial conditions (Table~\ref{tab:cases}) as well as the runtime properties are shown in Section~\ref{sec:results}, followed by a discussion of the findings in Section~\ref{sec:results} and finally a conclusion in Section~\ref{sec:Conclusion}.

\begin{table}[t]
\begin{ruledtabular}
\begin{tabular}{ccc}
\textrm{Case}&
\textrm{Temperatures}&
\textrm{Correlations}\\
\colrule
\includegraphics[width=0.25\linewidth, margin=0pt 1.5ex 0pt 0ex, valign=m]{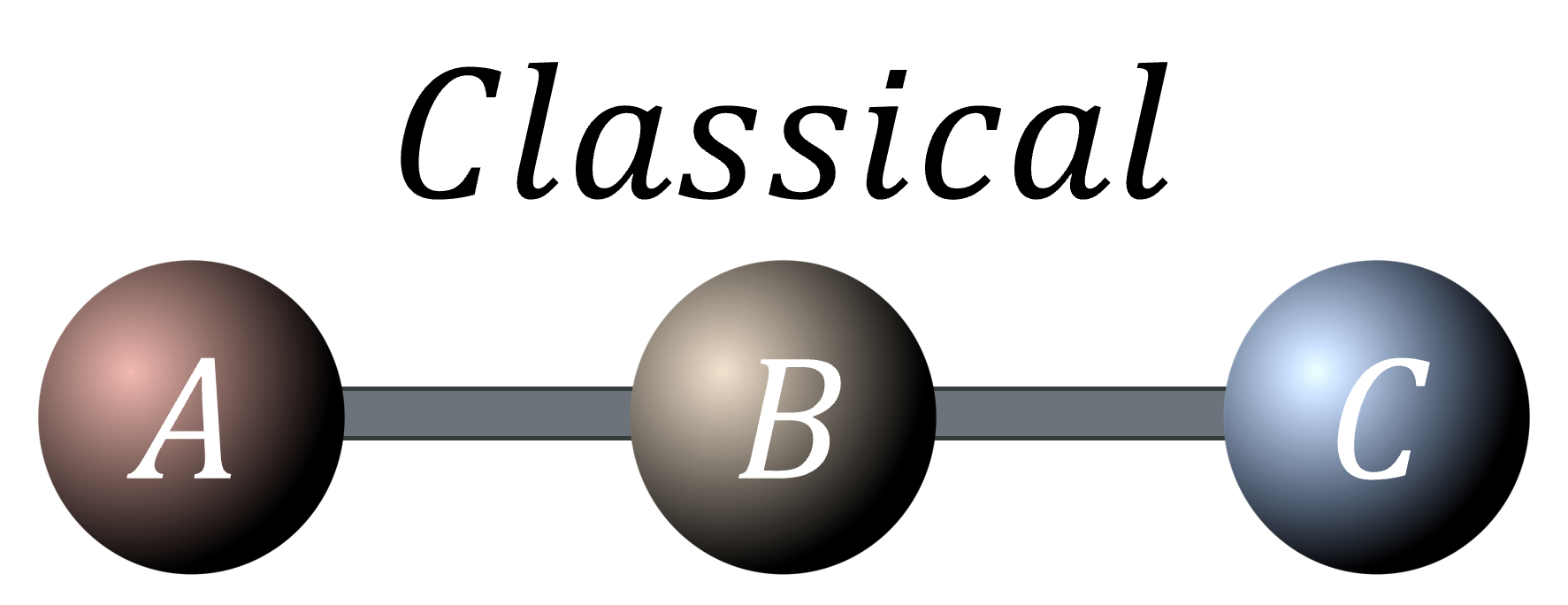} & $T_A > T_B > T_C$ & $\alpha_{AB} = \alpha_{BC} = 0$\\
\includegraphics[width=0.25\linewidth, margin=0pt 1.5ex 0pt 0ex,valign=m]{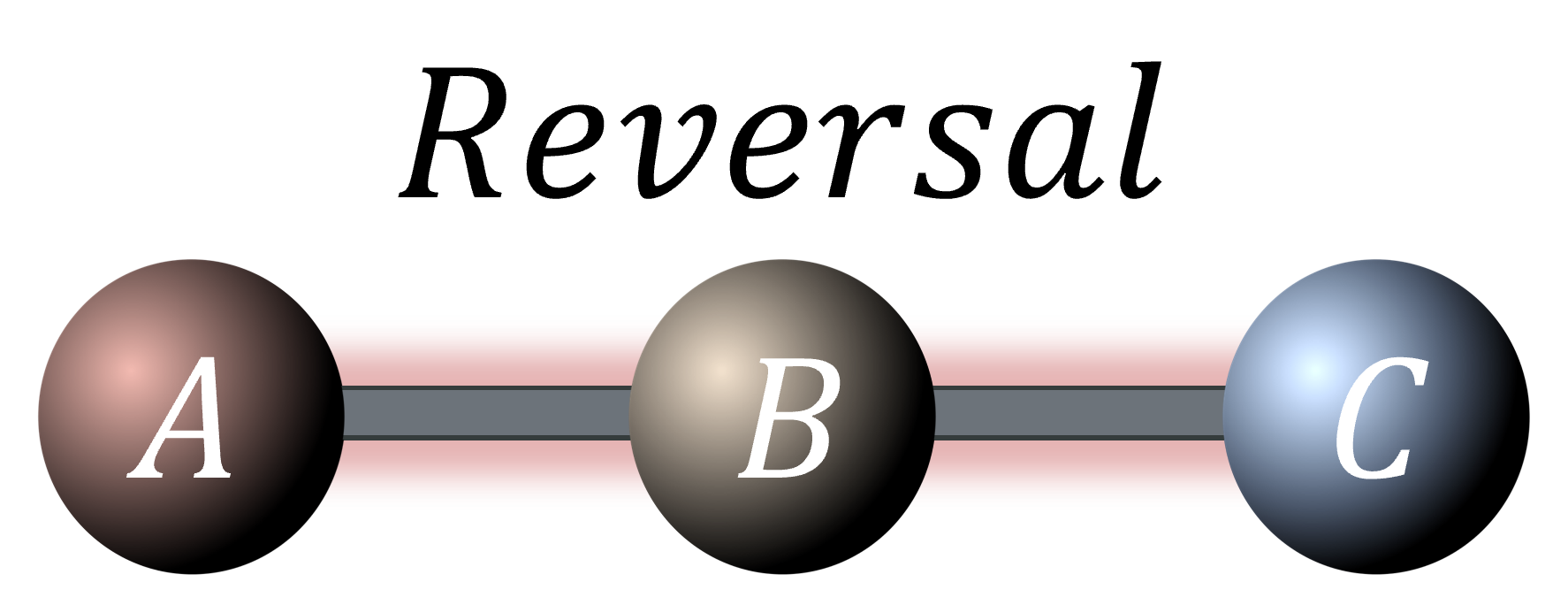} & $T_A > T_B > T_C$ & $\alpha_{AB},\,\alpha_{BC}<0$ \\
\includegraphics[width=0.25\linewidth, margin=0pt 1.5ex 0pt 0ex,valign=m]{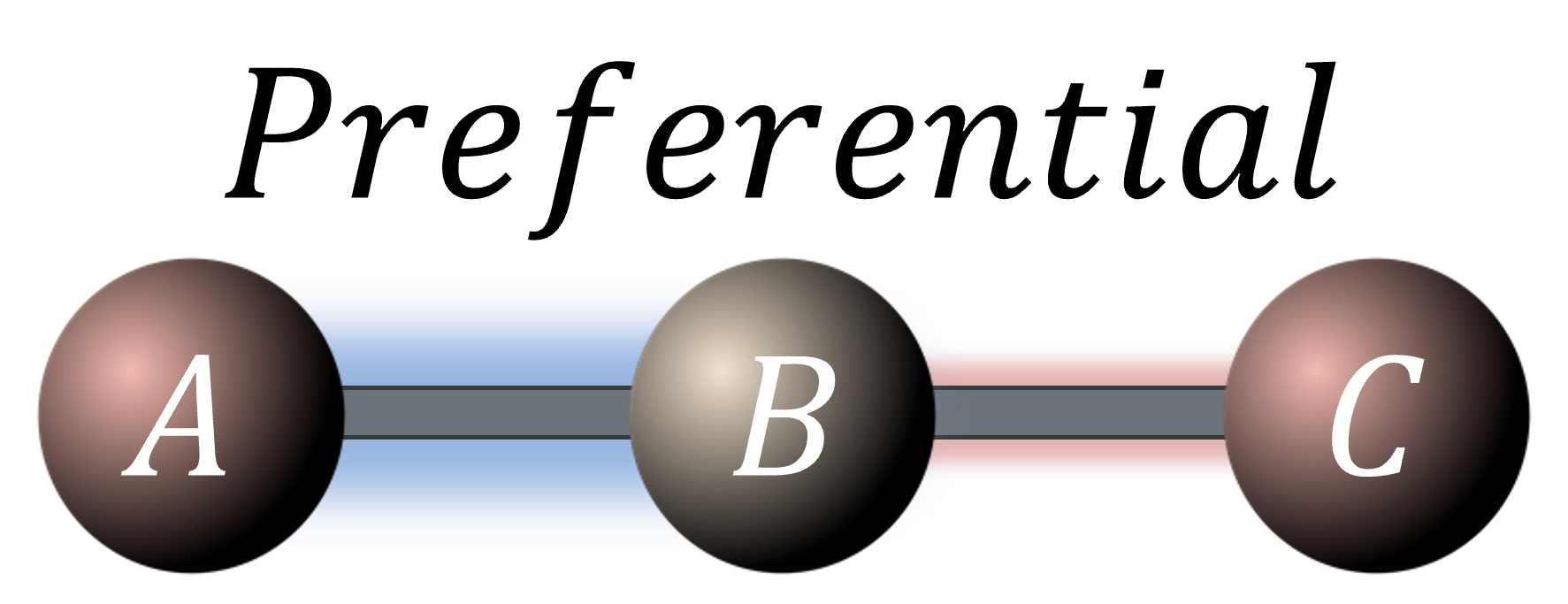} & $T_A = T_C$ & $-1< \frac{\alpha_{BC}}{\alpha_{AB}}<0$\\
\includegraphics[width=0.25\linewidth, margin=0pt 1.5ex 0pt 0ex,valign=m]{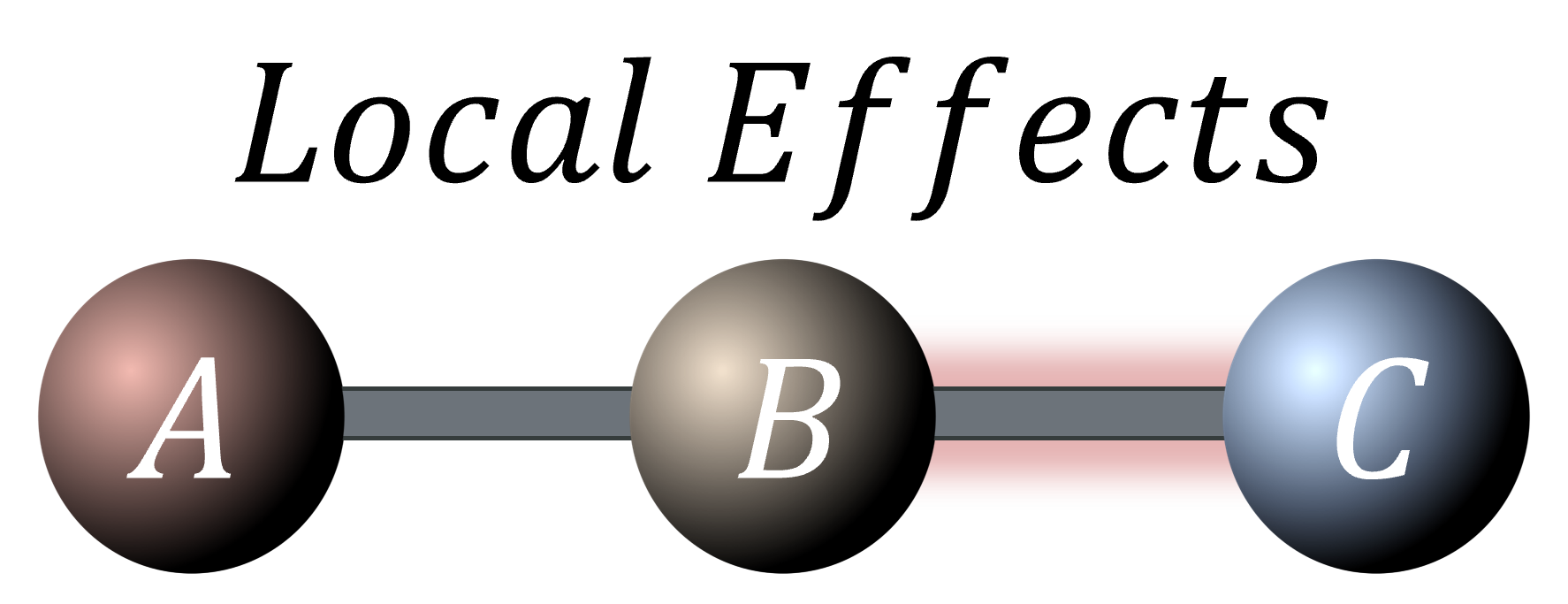} & $T_A > T_B > T_C$& $\alpha_{BC} < \alpha_{AB} = 0$\\
\end{tabular}
\end{ruledtabular}
\caption{\label{tab:cases}%
(Color online)
Four initial conditions of interest for the three-qubit chain as described by three initial temperatures and two initial coupling correlations. In descending order of rows, the cases follow: Classical, Reversal, Preferential Pumping, and Local Effects.}
\end{table}




\section{\label{sec:theory}Theory}
\subsection{\label{sec:bg}Quantum Thermodynamics Background} 
When adapted to a quantum state described by the density matrix $\rho$, the quantum analogue of the second law of thermodynamics reads
\begin{equation}
S(\rho_{\text{final}}||\rho_{\text{initial}})\geq 0\,,
\label{eq:q2ndlaw}
\end{equation}
where $S(\rho_{\text{final}}||\rho_{\text{initial}}) = -\Tr[\rho_{\text{final}}\log(\rho_{\text{initial}})] - \rho_{\text{final}}$ is the quantum relative entropy of the final state $\rho_{\text{final}}$ with respect to the initial state $\rho_{\text{initial}}$, and $S(\rho) = -\Tr[\rho \ln(\rho)]$ is the von Neumann entropy.

Consider two systems $i\in\{A, B\}$ that are each initially in a Gibbs state of the local Hamiltonian $\mathcal{H}_i$: 
\begin{equation}
\rho^i = \frac{e^{-\beta_i\mathcal{H}_i}}{\Tr[e^{-\beta_i\mathcal{H}_i}]},
\label{eq:gibbs}
\end{equation}
where $\beta = \frac{1}{k_{B}T}$ and $T$ is the temperature of the system. Provided that the evolution of the joint system $\rho^{AB} = \rho^A \otimes \rho^B$ is unitary and there is no work done throughout the evolution, it can be shown from \eqref{eq:q2ndlaw} and the properties of the quantum relative entropy that \cite{partovi,micadei}
\begin{subequations}\label{heatflow:main}
\begin{equation}\tag{\ref{heatflow:main}}
\begin{split}
\beta_B Q_B + \beta_A Q_A = Q_B(\beta_B - \beta_A) &\geq \Delta I(A:B).
\end{split}
\end{equation}
Here, $I(A:B) = S(\rho^A) + S(\rho^B) - S(\rho^{AB}) \geq 0$ is the mutual information which is a measure of the degree of correlation between the two systems. Note that the right-hand side of \eqref{heatflow:main} involves the \textit{change} in mutual information, which may be negative. 

Now consider the case when $A$ is hotter than $B$ ($\beta_A < \beta_B$). If $A$ and $B$ are initially uncorrelated, $\Delta I(A:B) \geq 0$. Thus
\begin{equation}
Q_B(\beta_B - \beta_A) \geq 0.
\label{heatflow:standard}
\end{equation}
It follows that $Q_B \geq 0$ and heat flows from $A$ to $B$, as is reminiscent of classical theory. In the presence of initial correlations, however, the mutual information can decrease $\Delta I(A:B) \leq 0$ so that the lower bound in \eqref{heatflow:main} is relaxed. It is now possible that
\begin{equation}
Q_B(\beta_B - \beta_A) \leq 0,
\label{heatflow:reversal}
\end{equation}
and heat can flow from the colder body to the hotter body, $Q_B \leq 0$. 
\end{subequations}

\subsection{\label{sec:setup}Physical Setup}
We now consider the system of a linear chain of $N$ spins along which only adjacent pairs interact. We subject the system to an external magnetic field in the $\bm{\hat{z}}$ direction so that the local Hamiltonian defining the initial state of each qubit as per Eq.\ \eqref{eq:gibbs} is 
\begin{equation}
\mathcal{H}^{i} = \frac{1}{2}h\nu_0(1-\sigma_z^i),
\label{eq:Hlarmor}
\end{equation}
where $\nu_0$ is the Larmor frequency \cite{micadei}.

In order to quantify the heat flow of each individual qubit along the chain, the state variables of each qubit such as its temperature and internal energy,
\begin{equation}
U^i = \Tr_i[\rho\mathcal{H}^{i}],
\label{eq:energycalc}
\end{equation}
must be well-defined. This introduces a constraint on our system whereby the local state of each qubit $\rho_i = \Tr_{i}[\rho]$ must remain a Gibbs state (refer to Eq. \eqref{eq:gibbs}) with respect to the Hamiltonian (defined in Eq. \eqref{eq:Hlarmor}) throughout the course of its evolution. We will call this constraint the Local Gibbs Criterion (LGC), which will be accounted for in the following characterisation of the initial state and evolution of our system. 
 
\subsubsection{\label{sec:isp}Initial State}
We generalise the definition of the initial state introduced by Meicadei et al.~\cite{micadei} for a chain of two qubits $i$ and $j$, 
\begin{equation}
    \rho_{\chi}^{i,\,j} = \rho_{0}^i \otimes \rho_{0}^j + \chi^{i,\,j},
    \label{ispgen:pair}
\end{equation}
to define the initial state
\begin{equation}
    \rho_0 = \sum_{i=1}^{N-1} \rho_0^{1\rightarrow i-1}\otimes \rho_{\chi}^{i,i+1}\otimes\rho_0^{i+2\rightarrow N} - (N-2)\rho_0^{1\rightarrow N},
    \label{ispgen:chain}
\end{equation}
for a chain of $N\geq3$ qubits, where the second term in \eqref{ispgen:chain} is incorporated to offset the double-counting of diagonal terms in the resultant mixed state, and $\rho_0^{i\rightarrow j} = \bigotimes_{k=i}^{j}\,\rho_0^k$ is adopted as shorthand notation. Notably, this initial state does \textit{not} incorporate correlations between the two endmost qubits which would otherwise wrap the chain into a ring.

 Crucially, this definition of the spin chain allows one to incorporate correlations for every adjacent pair of qubits. This is achieved by the inclusion of the $\chi^{i,\,j}$ term in Eq.~\eqref{ispgen:pair} which in this paper is chosen to be
\begin{equation}
\chi^{i,\,j} = 
\begin{bmatrix}
0 & 0 & 0 & 0\\
0 & 0 & \Bar{\alpha}_{i,\,j} & 0\\
0 & \alpha_{i,\,j} & 0 & 0\\
0 & 0 & 0 & 0\\
\end{bmatrix}
\label{eq:corrmatrix}
\end{equation}
and captures the initial correlations between qubits $i$ and $j$ \cite{micadei}. The parameter $\alpha_{i, j}$ is a measure for the correlations between the two qubits and its value is chosen to preserve the positivity of $\rho_0$. Since $\Tr_{i}[\chi_{AB}]=0$, it is ensured that introducing correlations of this form does not alter the condition that the qubits be in a local Gibbs state with respect to the Hamiltonian \eqref{eq:Hlarmor} and so have a well-defined internal energy \eqref{eq:energycalc} and temperature. Moreover, the reduced density matrix obtained by tracing out all but a given adjacent pair of qubits matches the form of \eqref{ispgen:pair} so that the correlation terms can be read off.

In total, there are $2N-1$ parameters defining the initial state of the chain: $N$ temperatures $\bm{T} = (T_1,\ldots,T_N)$ and $N-1$ nearest-neighbour correlations $\bm{\alpha} = (\alpha_1,\ldots,\alpha_N)$.

\subsubsection{\label{sec:evo}Thermalisation operator}
We recall from Refs.~\cite{micadei,jennings} that the thermalisation of two qubits can be described by a Dzyaloshinskii–Moriya (DM) interaction
\begin{subequations}\label{DM:main}
\begin{equation}
\mathcal{H}_{DM}^{i,j} = (h/2)J(\sigma_x^i\sigma_y^j - \sigma_y^j\sigma_x^i),
\label{DM:pair}
\end{equation}
where the frequency $J$ is a parameter describing the strength of interaction between qubits.  
For the $N$-qubit chain, we make the extension 
\begin{equation}
\mathcal{H}_{DM}= \sum_{i=1}^{N-1} \mathcal{H}_{DM}^{i,\,i+1},
\label{DM:chain}
\end{equation}
\end{subequations}
to allow for simultaneous thermalisation between all adjacent pairs, which yields the time evolution operator
\begin{equation}
U_\tau = e^{-i\tau \mathcal{H}_{DM}/\hbar}.
\label{eq:evogen}
\end{equation}

Importantly, since the thermalisation operator commutes with the sum of local Hamiltonians,
\begin{equation}
\biggl[\,\sum_{i=1}^{N}\mathcal{H}^i, \mathcal{H}_{DM}\biggr]=0\nonumber,
\end{equation}
thermalisation does not perform work on the total system. As in Ref.~\cite{jennings, micadei}, we assume that no mechanical work is done between the qubits so that change in mean energy of either system is due completely to the heat: $Q^i = \Delta U^i$.


Thus, heat flow may be inferred by calculating the energy of each qubit according to Eq.~\eqref{eq:energycalc} as a function of time. It should be noted that in general, evolving the initial state defined in Eq.~\eqref{ispgen:chain} according to the evolution operator in Eq.~\eqref{eq:evogen} also introduces correlations between non-neighbouring qubits. Compared to the correlations between adjacent qubits these are relatively weak.




\section{\label{sec:method}Circuit Implementation for the Case of Three Qubits}

\begin{figure*}[!ht]
\centering
\subfloat[\label{fig:layout}]{%
  \includegraphics[width=\columnwidth]{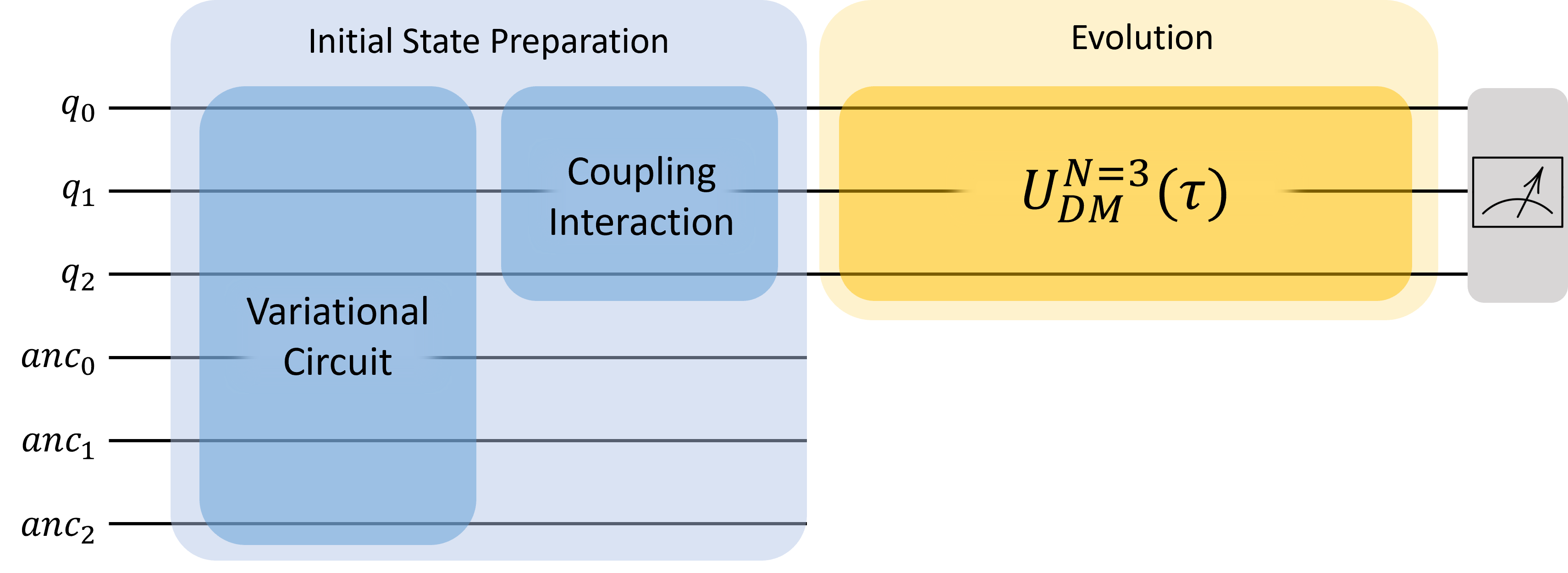}%
}\hfill
\subfloat[\label{fig:ansatz}]{%
\includegraphics[width=\columnwidth]{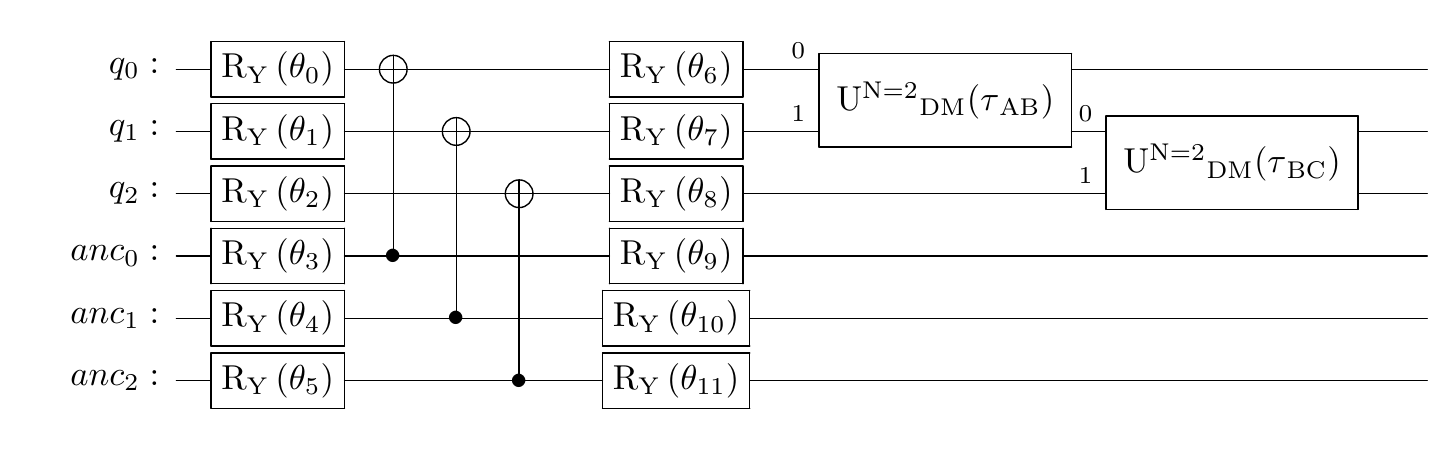}%
}

\subfloat[\label{fig:cartan}]{%
  \includegraphics[width=2\columnwidth]{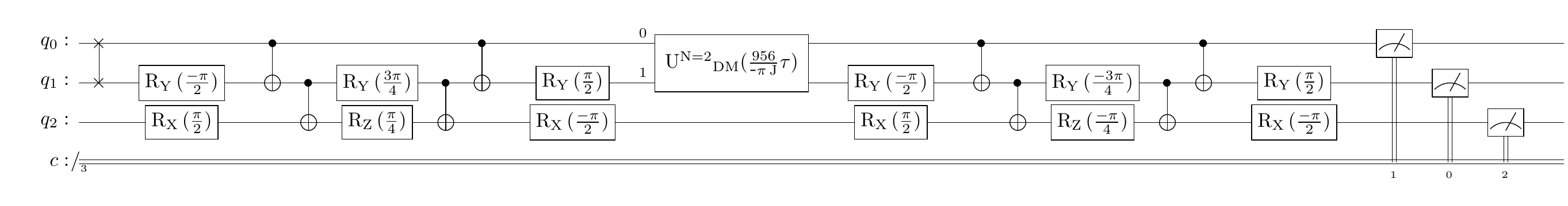}%
}\hfill
\subfloat[\label{fig:DM2}]{%
  \includegraphics[width=\columnwidth]{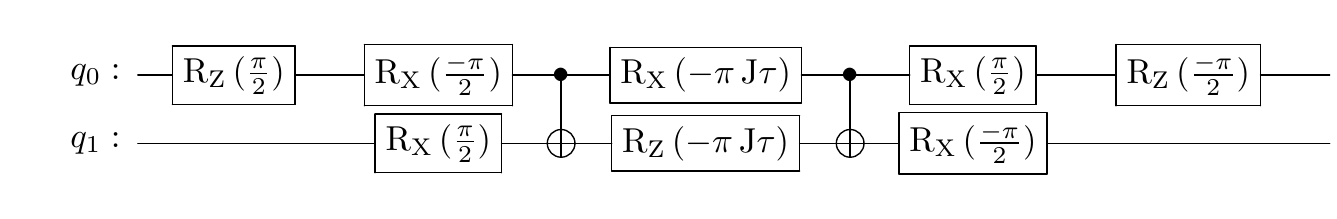}%
}
\caption{The relevant components of the circuit implementation for the three-qubit spin chain. (a) outlines the schematic layout of the circuit which consists of two stages: initial state preparation (b) and evolution (c). The initial state preparation (b) itself involves a \textit{variational} step for preparing an uncorrelated thermal state and a \textit{coupling} step for generating correlations. Note the label $anc$ for the bottom three qubits denotes the ancillary register, while the top three are system qubits. Below this is the two-qubit thermalisation subroutine $U^{N=2}_{DM}$ generated by the Hamiltonian \eqref{DM:pair}. The evolution stage (c) includes the Cartan circuit decomposition for $U^{N=3}_{DM}$ which makes use of $U^{N=2}_{DM}$ (d); optimisations made to the conjugate terms come at the expense of an initial swap gate whose effect is reversed during measurement by redirecting qubit correspondence to the classical register (see Appendix \ref{sec:cartan} for more information)\label{fig:circuits}}
\end{figure*}

We now use a quantum computer to simulate the heat flow along the three-qubit chain where $N=3$ and $i\in\{A,B,C\}$.   Table~\ref{tab:cases} shows four different instances of theoretical interest: 
\begin{description}[before={\renewcommand\makelabel[1]{\bfseries ##1}}]
  \item[Classical] \hfill \\
  The classical case contains a thermal gradient $A\rightarrow\,C$ but without initial correlations existing between any adjacent pair of qubits. As a result, we expect the initial heat flow to be in the direction of the temperature gradient as would be predicted by classical thermalisation. 
  \item[Reversal] \hfill \\ 
  The reversal case contains a similar temperature gradient to the classical case, except that now there exist nonzero initial correlations between both adjacent pairs of qubits, $\alpha_{AB}$ and $\alpha_{BC}$. This generalises the original case of two correlated qubits demonstrated in Ref.~\onlinecite{micadei} so we expect to see an analogous reversal of heat against the temperature gradient.
  \item[Preferential pumping] \hfill \\ In this case, we explore the effect of the magnitude of initial correlations on heat flow. To achieve this, we set the temperatures of the two endmost qubits $T_{A}= T_{C}$ to be the same and hotter than the middle qubit, while enforcing stronger correlations between $A$ and $B$, than $B$ and $C$. Ignoring correlations -- as one would in a classical regime -- we expect the heat flow to be symmetric with the same amount of heat leaving both $A$ and $C$ to be absorbed by $B$. In contrast, in the correlated case we expect there to be a preference of heat flow from the middle spin to to one of the endmost spins.
  \item[Local effects] \hfill \\ This case follows almost the same set-up as the reversal case, except we remove correlations between one of the adjacent pairs so that $\alpha_{AB} = 0$. Here we probe how a local heat reversal between $B$ and $C$ might influence the heat flow to and from $A$ which
    initially is not correlated with the rest of the system and thus expected to behave classically.
\end{description}

In line with the layout presented in Section~\ref{sec:setup}, the schematic layout of our circuit implementation shown in Fig.~\ref{fig:layout} consists of two routines: initial state preparation (Fig.~\ref{fig:ansatz}), and evolution (Fig.~\ref{fig:cartan}). The fixing of parameters $\mathbf{\tau}$ and $\mathbf{\theta}$ in both these routines relies on a posteriori data obtained from classical simulations which was made possible by the sufficiently small size of the system. 

A common subroutine that will be used in what follows is the coupling operator defined in Eq.~\eqref{eq:evogen} for $N=2$, the circuit implementation of which can be seen in Fig.~\ref{fig:DM2}.


\begin{figure*}[!ht]
\subfloat[\label{fig:a}]{%
  \includegraphics[width=\columnwidth]{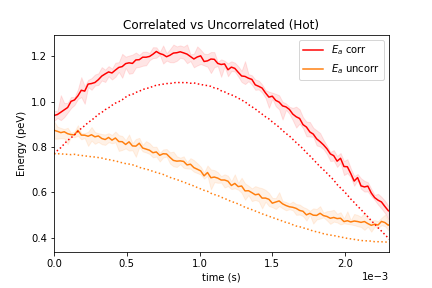}%
}\hfill
\subfloat[\label{fig:b}]{%
  \includegraphics[width=\columnwidth]{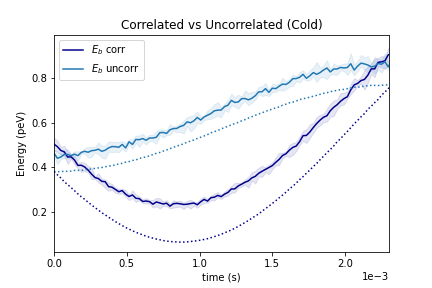}%
}

\subfloat[\label{fig:c}]{%
  \includegraphics[width=\columnwidth]{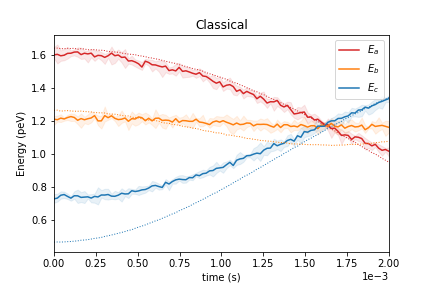}%
}\hfill
\subfloat[\label{fig:d}]{%
  \includegraphics[width=\columnwidth]{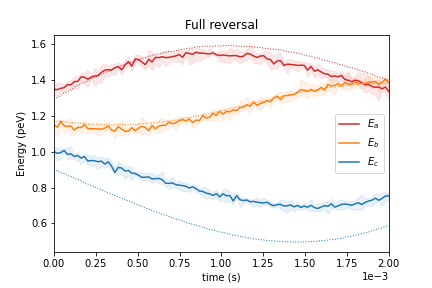}%
}

\subfloat[\label{fig:e}]{%
  \includegraphics[width=\columnwidth]{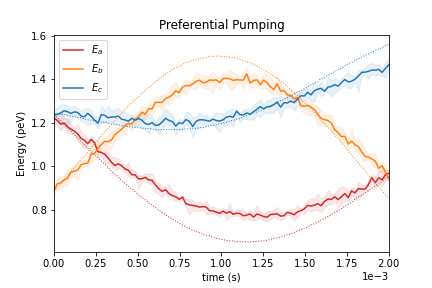}%
}\hfill
\subfloat[\label{fig:f}]{%
  \includegraphics[width=\columnwidth]{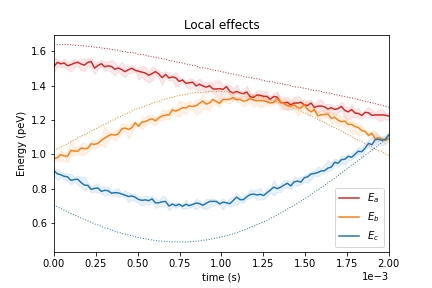}%
}

\caption{The reproduced heat flow for the two-qubit case \cite{micadei} consisting of a hot (a) and cold (b)  qubit. (c)-(f) display the heat flow for the three-qubit case under the different instances mentioned in Table~\ref{tab:cases} including classical (c), reversal (d), preferential pumping (e), and local effects (f). Results were obtained by running on the Qiskit \textit{ibmq}\_\textit{toronto} backend,  averaged over 6 trials each consisting of 8192 shots. Shown in dotted are the simulation predictions. }\label{fig:results}

\end{figure*}

\subsubsection{\label{sec:isp_method}Initial State Preparation}
We now aim to prepare the initial state given in Eq.~\eqref{ispgen:chain}. Our preparation strategy is to create an uncorrelated chain first, onto which we then imprint the desired correlations.

\paragraph{Uncorrelated state preparation}
An uncorrelated state corresponds to a diagonal density matrix in which all off-diagonal terms are zero. For the definition introduced in Eq.~\eqref{ispgen:chain}, this corresponds to setting the correlation terms $\bm{\alpha} = \bm{0}$. Such a state may be prepared using a variational circuit (Fig.~\ref{fig:ansatz}) which is parameterised by twelve Y-rotation angles $\bm{\theta}$ and entangles every qubit in the chain with a respective ancilla qubit. As a result, the number of qubits in our register totals $2N = 6$ and the state $\rho'_{uncorr}$ of the chain is recovered by ignoring the ancillary register. 

We then use a classical optimiser (viz.\  Gradient descent) with a Manhattan distance loss function to obtain the rotation angles that best approximate $\rho'_{uncorr}$ to a provided $\rho_{uncorr}$. 

\paragraph{Generating correlation terms}




The general form of Eq.~\eqref{ispgen:chain} can be approximated by applying two consecutive two-qubit operators (Figure~\ref{fig:ansatz}) parametrised by propagation times $\tau_{AB}$ and $\tau_{BC}$ to each of the two pairs $(A,B)$ and $(B,C)$ respectively, obtaining $\rho'_{corr}$.

When this construction (Fig.~\ref{fig:ansatz}) is applied to a diagonal state for arbitrary propagation times, the resultant $\rho'_{corr}$ generally also has extraneous off-diagonal terms giving rise to a nonzero correlation $\alpha_{AC}$, wrapping our chain into a ring. Nonetheless, in all the four cases investigated, the magnitude of $\alpha_{AC}$ remains small compared to $\alpha_{AB}$ and $\alpha_{BC}$ (see Appendix). For example, in one experiment, $(\alpha_{AB}, \alpha_{BC}, \alpha_{AC}) = (-0.097, -0.076, -0.012)$.

The goal now becomes tuning $\bm{\tau} = (\tau_{AB}, \tau_{BC})$ to obtain the cases of interest in Table~\ref{tab:cases}. We obtain $\bm{\alpha}(\bm{\tau})$ and $\bm{T}(\bm{\tau})$ from $\rho'_{corr}$. Note that  $\bm{\alpha}(\bm{0}) = \bm{0}$ yields the classical case discussed in Section \ref{sec:method}
so long as the uncorrelated state prepared is one in which $\bm{T}(\bm{0}) = (T_A, T_B, T_C)$ satisfies $T_A > T_B > T_C$. We may now choose a relevant cost function with arguments $\bm{\alpha}(\bm{\tau})$ and $\bm{T}(\bm{\tau})$ that when optimised using a classical optimiser returns a solution that
satisfies the conditions of the case under discussion.


\begin{figure*}[!ht]
\subfloat[\label{fig:var}]{%
  \includegraphics[width=\columnwidth]{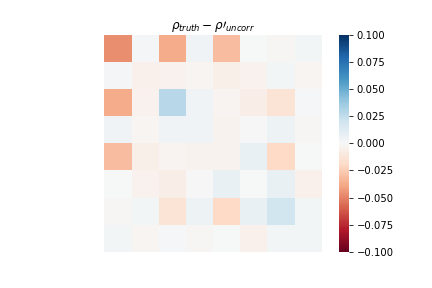}%
}\hfill
\subfloat[\label{fig:pair}]{%
\includegraphics[width=\columnwidth]{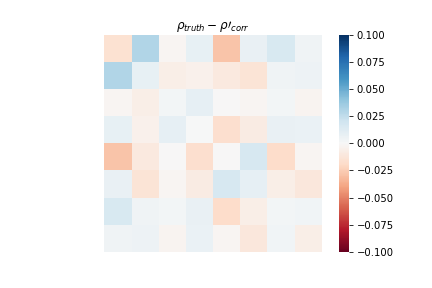}%
}

\subfloat[\label{fig:all}]{%
  \includegraphics[width=\columnwidth]{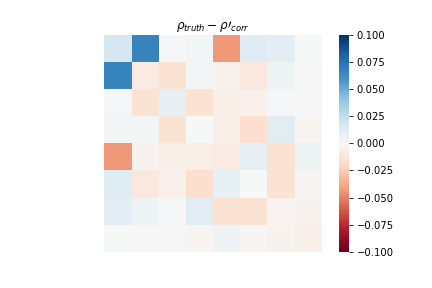}%
}\hfill
\subfloat[\label{fig:evo}]{%
  \includegraphics[width=\columnwidth]{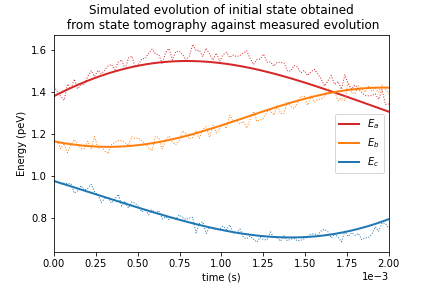}%
}
\caption{The error in the measured state at $\tau=0$ obtained by state tomography for the reversal case compared to the theoretical state at three different stages along the final circuit implementation: variational (a), coupling (b), and evolution (c). The errors emerge as a result of unwanted qubit thermal relaxation and rotation. (d) shows the correspondence between the heat flow obtained by measurement, and that obtained by numerically evolving the measured initial state obtained by state tomography.}\label{fig:main}
\end{figure*}

\subsubsection{\label{sec:evo_method}Thermalisation operator}
One obvious circuit implementation for
$U^{N=3}_{DM}(\tau)$ is a
Suzuki–Trotter approximation which would involve repeated application of the coupling interaction shown in the second half of the circuit in Figure \ref{fig:ansatz}.
However, if a low margin of error is required, this method quickly becomes costly in terms of circuit depth and particularly the number of CNOT gates. 
Alternatively, to  obtain a relatively inexpensive and fixed-depth implementation we employed a Cartan decomposition  \cite{cartan} which requires 18 CNOTs and makes use of $U^{N=2}_{DM}(\tau)$. 
A further reduction to 13 CNOTs can be made by condensing the conjugate terms involved in the decomposition to yield the circuit in Fig.~\ref{fig:cartan} (see Appendix \ref{sec:cartan}).

Our key goal in the design of the circuits is minimising the circuit depth so as to reduce the noise in the energy readings. In particular, we wish to minimise the CNOT count and avoid the use of SWAP gates by applying CNOTs only between qubits that are neighbouring with respect to the coupling map of the IBM quantum machine being used.

It should be noted that these technical accommodations are due to the extension from a two- to three-qubit chain as the terms in the DM interaction for three qubits no longer commute. 




\section{\label{sec:results}Results and Discussion}

Fig.~\ref{fig:results} displays the heat flows for both the reproduced two-qubit case \cite{micadei}, as well as the four instances of the extended three-qubit case. The results were obtained by running on the 27-qubit IBM quantum backend \textit{ibmq}\_\textit{toronto}, where measurement error mitigation was used to reduce readout errors. The coupling map of this backend provided several choices for an optimal group of six physical qubits on which to run the circuit implementation for the three-qubit chain \footnote{Optimal in the sense of not requiring any additional SWAP gates to accommodate CNOTs between any pair of qubits along the circuit}. The subset of qubits chosen was that which yielded initial energy measurements at $\tau=0$ which were closest to the theoretical values. It was assumed that the subset which met this criterion would also offer results that were closest to the overall theoretical heat flow curve for all other times. In measuring the non-classicality of the initial correlations the Geometric Quantum Discord (GQD) \cite{gqd} was calculated for each pair of qubits (using the initial density matrix as obtained by state tomography) and then compared to the theoretical values.

Figs.~\ref{fig:a} and \ref{fig:b} showcase the heat flow between two qubits comparable to the findings of Micadei et al.\ for two qubits using an NMR setup \cite{micadei}. The energy of the hotter qubit shown in  Fig.~\ref{fig:a} initially increases (decreases) in the correlated (uncorrelated) case, highlighting heat reversal as brought upon by introducing quantum correlations. The opposite effect occurs for the energy of the colder qubit shown in Fig.~\ref{fig:b}.

In contrast, Figs.~\ref{fig:c} and \ref{fig:d} show the difference in heat flow between the classical and correlated cases respectively for our extended three-qubit system. The GQDs between the pairs A \& B, and B \& C, in the initial state of the correlated case are $2.3\times10^{-2}$ and $1.7\times10^{-2}$ respectively, significantly higher than the corresponding values of $1\times10^{-3}$ and $8\times10^{-3}$ obtained in the uncorrelated case. As a result, while in Fig.~\ref{fig:c} heat is shown to flow classically along the temperature gradient $A\rightarrow\,B\rightarrow\,C$, introducing correlations $\alpha_{AB}$ and $\alpha_{BC}$ induces a full reversal of the heat flow in the direction $C\rightarrow\,B\rightarrow\,A$ as seen in Fig.~\ref{fig:d}. The times at which the hottest and coldest qubits each obtain their energy extrema appear to be out of phase, unlike in the two-qubit case, which highlights the role of the intermediate qubit both as a conduit as well as a storage of heat. 

It should be noted that while, theoretically, it would be expected for the GQD between any pair of qubits to evaluate to zero in the uncorrelated case, the fact that it is still nonzero indicates the presence of noise introducing some amount of quantum correlations. Indeed, even running an empty circuit produces a state that is similarly quantum correlated -- if only weakly so -- which implies that even the initial state is not necessarily $\ket{\Tilde{0}}$. It must also be mentioned that the density matrix used in the GQD calculations was obtained by state tomography whose circuit implementation is equally prone to noise. 

In Fig.~\ref{fig:e}, we observe more clearly the role of quantum correlations as a heat pump. By having $|\alpha_{AB}| > |\alpha_{BC}|$, the GQD between $A$ and $B$ ($2.3\times10^{-2}$) evaluates to be almost an order of magnitude larger than the GQD between $B$ and $C$ ($4\times10^{-3}$). As a result, despite there being a symmetric temperature distribution, qubit $A$ is the preferred heat source for the central qubit $B$, even though both $A$ and $C$ are initially at the same temperature. 

Finally, by enforcing relatively large correlations between only one adjacent pair, it is possible to restrict heat reversal only to that pair alone. This is seen in Fig.~\ref{fig:e} for the pair $B$ and $C$ (GQD: $1.5\times10^{-2}$) compared to the pair $A$ and $B$ (GQD: $4\times10^{-3}$), as qubit $A$ continues to thermalise classically, albeit at a slower rate due to the anomalous heating of its neighbour $B$ which is correlated with $C$. This reveals a property of locality in the proposed generalisation, in that the heat reversal appears to only occur between neighbours that share a correlation.

The slight inconsistency in the results for the three-qubit case against the simulated predictions may be explained by the presence of errors in as early as the variational step which despite being intended for preparing an uncorrelated and diagonal state still exhibits off-diagonal terms as shown in the heatmap of Fig.~\ref{fig:var}.
In Fig.~\ref{fig:pair} we see that these errors are dispersed throughout the mixture during the coupling stage of initial state preparation,
and then intensified after the evolution stage Fig.~\ref{fig:all}, which for the case of $\tau=0$ has the benign effect of only increasing the circuit depth, and thus, the chance for errors to propagate. This is further emphasised by the close correspondence between numerically evolving the state obtained via state tomography and the real results in Fig.~\ref{fig:evo} which suggests that the evolution stage is not the main source of error. In addition to this, the systematic overestimation of the real trials compared to the predicted results for the two-qubit case in Figs.~\ref{fig:a} and \ref{fig:b} suggests a thermal relaxation error caused by short thermal relaxation time $T_1$.

\section{\label{sec:Conclusion}Conclusion}
We have simulated anomalous heat flow for the case of a three-qubit spin chain on a quantum computer. Our circuit implementation prepares and evolves the state of the chain, and yields reliable results on noisy quantum computers. This extends previous results obtained by Micadei et al.\ for the smallest case of a two-qubit chain \cite{micadei}.

In contrast to classical thermalisation, we showcased a range of exotic behaviours by investigating four main initial conditions for quantum correlations. These included the full and partial reversal of heat which provides further experimental verification of the nature of decreasing quantum correlations as an entropic process which can offset the entropy consumed during heat reversal \cite{partovi}.
In addition to this, we observed the asymmetric pumping of heat in an otherwise symmetric temperature distribution, hinting at the utility of correlations as a means of precisely controlling heat in small systems.

One may consider a circuit implementation for a spin chain of arbitrary length by following a similar procedure to the special case of the three-qubit chain. However, extending this to large lengths is currently limited by the a posteriori simulation data that is used to fix the parameters for initial state preparation. As far as uncorrelated state preparation is concerned, a priori methods have been devised for preparing Gibbs states \cite{varqite}, but despite these the task of generating correlations remains nontrivial. Simulating longer chains will respectively require more gates and a longer circuit depth, worsening the effects of noise when run on current noisy machines.

One point of interest for future work is experimental demonstration of anomalous heat flow in systems more complex than a linear chain of qubits where additional exotic effects may emerge, such as a system of qudits. A possible extension to the setup may involve making the system non-isolated by having it interact with an external thermal bath such as the one defined in Ref.~\onlinecite{bath}. This would allow closer modelling of the dissipation of heat which is more physically accurate than assuming that the system is closed and no heat is lost. Further, initial correlations may be utilised in this setting to achieve more ambitious thermodynamic goals, such as thermally insulating the system from the bath.
\begin{acknowledgments}
This work was supported by the University of Melbourne through the establishment of an IBM Quantum Network Hub at the university. C.D.H.\ is supported through a Laby Foundation grant at the University of Melbourne. T.Q.'s research was conducted at the Australian Research Council (ARC) Centre of Excellence for Mathematical and Statistical Frontiers (ACEMS, project number CE140100049) and partially funded by the Australian Government.
\end{acknowledgments}

\begin{figure*}
  \begin{captivy}{\includegraphics[scale=0.9]{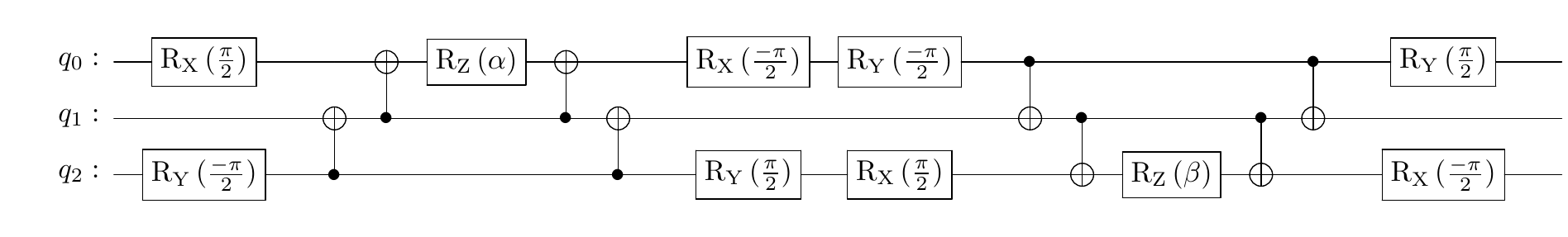}}
    \oversubcaption{0, 0.8}{}{fig:18cnot}
  \end{captivy}
  \begin{captivy}{\includegraphics[scale=0.9]{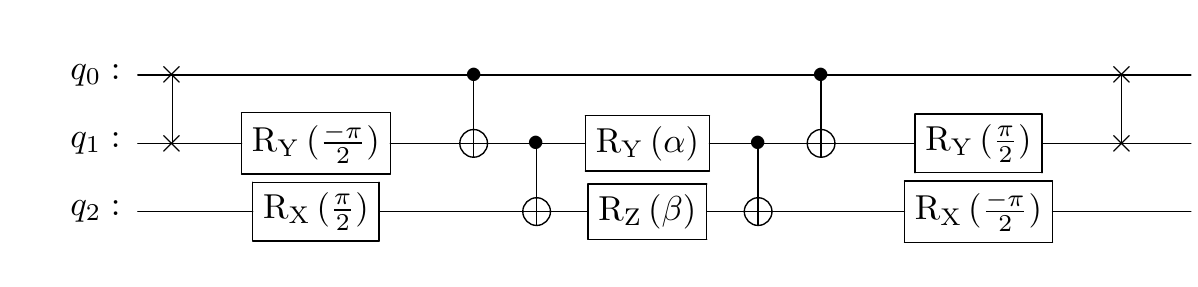}}
    \oversubcaption{0, 0.8}{}{fig:13cnot}
  \end{captivy}
  \caption{Two circuits for $K$ with forms $e^{-\frac{i}{2}\alpha XZY}e^{-\frac{i}{2}\beta YZX}$ (a) and $\big(I \otimes SWAP\big)\,e^{-\frac{i}{2}(\alpha XYZ + \beta YXZ)}\,\big(I \otimes SWAP\big)$ (b). (a) and (b) respectively require 8 and 10 CNOTs, and yield circuits for $U^{N=3}_{DM}$ which require 18 and 13 CNOTs, the latter of which is shown in Figure \ref{fig:cartan}.
\vspace{5mm}}
  \label{fig:optim}
\end{figure*}

\begin{table*}[!t]
\begin{ruledtabular}
\begin{tabular}{ccccccccccccc}
Case&\multicolumn{2}{c}{$\bm{\tau}$}&\multicolumn{3}{c}{$\bm{T}(\bm{\tau})$ (peV)}
&\multicolumn{3}{c}{$\bm{\alpha}(\bm{\tau})$} &\multicolumn{3}{c}{$\bm{D}(\bm{\tau})$}\\
&$\tau_{AB}$&$\tau_{BC}$&$T_A$& $T_B$& $T_C$& $\alpha_{AB}$
& $\alpha_{BC}$ & $\alpha_{AC}$ & $D_{AB}$ & $D_{BC}$ & $D_{AC}$ \\
\hline
Classical & 0 & 0 & 9.8 & 5.0 & 2.00 & 0 & 0 & 0 & 0 & 0 & 0\\
Full Reversal &
$-1.23\mathrm{e}{-3}$ & $-1.05\mathrm{e}{-3}$ & 5.3 & 4.5 & 3.2 & −0.097 & −0.071 & −0.014 & 0.037 & 0.021 & 0.001\\
Preferential Pumping &
$-8.78\mathrm{e}{+1}$ & $9.33\mathrm{e}{-4}$& 4.9 & 3.2 & 4.9 & 0.13 & -0.025 & 0. & 0.071 & 0.002 & 0\\
Local Effects &
$-1.01$ & $0$ & 9.9 & 3.7 & 2.6 & -0.089 & 0 & 0 & -0.031 & 0 & 0\\
\end{tabular}
\end{ruledtabular}
\caption{\label{tab:table2}
The parameters that define each of the four cases explored in the main paper. From left to right, the columns are the following: $\bm{\tau}$ is the time to which the coupling circuits are propagated in order to yield the required correlations; $\bm{T}(\bm{\tau})$ is the initial temperature of the qubits just before evolution; $\bm{\alpha}(\bm{\tau})$ is the initial correlation of the qubit pairs, including the extraneous term $\alpha_{AC}$ which is consistently small relative to the others for all cases; and $\bm{D}(\bm{\tau})$ is the Geometric Quantum Discord which is a measure of how quantum the correlations are in origin. Note that these valued are obtained through a simulator.}
\end{table*}

\bibliography{apssamp}
\newpage

\appendix
\section{Case Parameters}
Table \ref{tab:table2} tabulates the relevant parameters for the circuit implementation of each of the four cases.

\section{\label{sec:cartan}Optimising the circuit implementation of \texorpdfstring{$U^{N=3}_{DM}$}{}}
The general Cartan decomposition for the three-qubit thermalisation operator reads,
\begin{equation}
    U^{N=3}_{DM}(\tau) = K^\dagger\Big(I \otimes U^{N=2}_{DM}(c\tau)\Big)K\,,
    \label{eq:cartandesc}
\end{equation}
where $K = e^{-\frac{i}{2}(\alpha XZY + \beta YZX)}$ and constants $\alpha$,  $\beta$ and $c$ are obtained during the decomposition procedure \cite{cartan}.

Since the circuit for $U^{N=2}_{DM}$ is already known, all that remains is to find a circuit implementation for $K$. The two terms in the argument of the exponential commute, $[XZY, YZX] = 0$, so $K$ is separable as a product of Pauli string exponentials whose standard representation requiring 8 CNOTs is shown in \ref{fig:18cnot}. The total number of CNOTs required to implement $U^{N=3}_{DM}$ with this method is $(2\times8) + 2 = 18$. 

To reduce as much noise as possible it is necessary to minimise the number of CNOTs. This must be done under the constraints of the qubit coupling map of the respective backend which specifies the pairs of qubits  which can have a direct CNOT applied between them without requiring an auxiliary SWAP gate. In the case of $ibmq\_toronto$, our three qubits share a linear coupling so that qubits $q_0$ and $q_2$ are uncoupled.

With this constraint in mind, an alternative circuit implementation is shown in Figure \ref{fig:13cnot}. While the circuit itself requires 10 CNOTs and is more costly than the previous method, the implementation of $U^{N=3}_{DM}$ using this routine ends up consuming only $3 + (2\times4) + 2 = 13$ CNOTs as three of the four total SWAPs can be eliminated. This is because the left SWAP gate of $K^\dagger$ commutes with $U^{N=2}_{DM}$\footnote{that is, $U^{N=2}_{DM}$ is symmetric under interchange of qubits} and undoes the effect of the right SWAP gate of $K$. Moreover, the right SWAP of $K^\dagger$ -- as the last gate in the circuit -- can instead be incorporated into the measurement operator by simply swapping the qubit-to-bit correspondence as shown in Figure \ref{fig:cartan}.

\end{document}